\documentclass[useAMS,usenatbib]{mn2e}
\usepackage{graphicx}

\title[OH \& methanol masers in ON1.]{A MERLIN Study of 6 GHz Excited-state OH \& 6.7 GHz Methanol Masers in ON1}
\author[Green et al.]
{J. A. Green$^1$\thanks{E-mail:james.green@postgrad.manchester.ac.uk}, 
A. M. S. Richards$^1$, 
W. H. T. Vlemmings$^{1, 2}$, \newauthor
P. Diamond$^{1}$ 
and R. J. Cohen$^1$\thanks{Deceased 2006.} 
\\
$^1$ University of Manchester, Jodrell Bank Observatory, Macclesfield, Cheshire, SK11 1DL;\\
$^2$ Argelander Institute for Astronomy, University of Bonn, Auf dem H\"ugel 71, 53121 Bonn, Germany;
\\}
\begin{document}

\date{Accepted 2007 September 3. Received 2007 August 31; in original form 2007 May 31}

\pagerange{\pageref{firstpage}--\pageref{lastpage}} \pubyear{2007}

\maketitle

\label{firstpage}

\begin{abstract}
MERLIN observations of 6.668-GHz methanol and both 6.031- and
6.035-GHz hydroxyl (OH) emission from the massive star-formation
region ON1 are presented. These are the first methanol observations
made in full polarization using 5 antennas of MERLIN, giving high
resolution and sensitivity to extended emission. Maser features are
found to lie at the southern edge of the ultra-compact HII region,
following the known distribution of ground-state OH masers. The masers
cover a region $\sim$1 arcsec in extent, lying perpendicular to the
H$^{13}$CO$^{+}$ bipolar outflow. Excited-state OH emission
demonstrates consistent polarization angles across the strongest
linearly polarized features which are parallel to the overall
distribution. The linear polarizations vary between 10.0 and 18.5 per
cent, with an average polarization angle of
-60$^{\circ}$$\pm$28$^{\circ}$. The strongest 6.668-GHz methanol
features provide an upper limit to linear polarization of $\sim$1 per
cent.  Zeeman splitting of OH shows magnetic fields between
$-$1.1 to $-$5.8 mG, and a tentative methanol magnetic field strength
of $-$18 mG is measured.
\end{abstract}


\section[]{INTRODUCTION}
High-mass star-formation regions have been observed to demonstrate
maser emission from a wide variety of molecules, including both
methanol and hydroxyl (OH). One such region, ON1, is an ultra-compact
HII region (UCHII) contained within the Onsala molecular cloud. The
region was first observed by Elld\'er, R\"onn\"ang \& Winnberg in 1969
through its OH maser emission, before it was observed in CO, H$_{2}$CO
and HCO$^{+}$ in 1983 by Israel \& Wootten and shown to be part of an
extended molecular cloud complex. NH$_{3}$ and H76$\alpha$
recombination line observations led Zheng et al. (1985) to identify
ON1 as a rapidly rotating condensation, but Kumar, Tafalla \& Bachiller (2004) later
identified two H$^{13}$CO$^{+}$ outflows, demonstrating a resolved
bipolar structure with a velocity gradient of $\sim$30 km s$^{-1}$
pc$^{-1}$. ON1 is associated with the IRAS source 20081+3122. It is
believed that ON1 contains a central binary massive protostar of type
B0.3 surrounded by a young stellar cluster (Israel \& Wootten 1983;
Macleod et al. 1998; Kumar et al. 2004).

The kinematic distance estimate to ON1 is between $\sim$1.4 to 8 kpc
depending on the model used for the Galactic rotation (Israel \&
Wootten 1983; Kurtz, Churchwell \& Wood 1994; Macleod et al. 1998).
The currently favoured distance is the near kinematic distance of 1.8
kpc, as derived by Macleod et al. (1998), using the Wouterloot \&
Brand (1989) rotation curve. Macleod et al. (1998) estimate the age of
ON1 to be 0.5 $\times$ 10$^{5}$ years.

\begin{figure}
 \centering
\includegraphics[width=8cm]{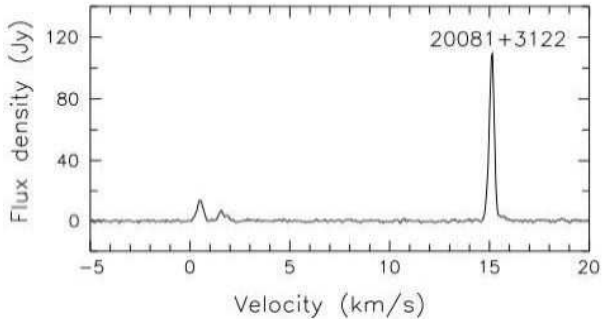}
\caption{\small Single-dish spectrum of 6.7-GHz methanol emission from
ON1 taken with the 32-m Torun telescope (Szymczak et al. 2000).}
\label{figure1}
\end{figure}

The 6.7-GHz methanol maser has an intrinsic relationship with
high-mass star-formation, having to date been observed in exclusive
association with known massive star-formation regions (Minier et
al. 2003).  Methanol maser emission at 6668.512 MHz was first observed
in ON1 with the 43-m Greenbank telescope in 1991 by Menten, who
detected a feature with a peak flux density of 91 Jy at +15.1 km
s$^{-1}$. This was later observed by Szymczak, Hrynek \& Kus (2000) and found
to have a peak flux density of 109 Jy. The spectrum of this
observation is given in Fig. \ref{figure1}, and reveals a weak feature
at around 0 km s$^{-1}$ in addition to the peak feature at +15.1 km
s$^{-1}$.

6.7-GHz methanol maser polarization has only been studied twice
previously, Ellingsen (2002) found fractional polarizations between a
few and 10 per cent in NGC6334F, whilst Vlemmings, Harvey-Smith \&
Cohen (2006a) found approximately 2 per cent, with levels up to 8 per
cent, in W3(OH). The methanol molecule is diamagnetic and as such has
low magnetic permeability and a small magnetic dipole moment and so
methanol masers in the presence of an external magnetic field are not
expected to demonstrate strong linear polarization, nor to show large
fractional circular polarization.

As the OH molecule is paramagnetic, OH masers are good tracers of the
magnetic field.  Ground-state OH towards ON1 was observed with MERLIN
by Nammahachak et al. (2006). The 6-GHz excited-state of OH is also
particularly effective for measuring the Zeeman effect (Caswell \&
Vaile 1995). Desmurs \& Baudry (1998) conducted VLBI observations of
6.035-GHz OH in ON1 in 1994, finding 4 right hand circularly polarized
(RHC) features and 3 left hand circularly polarized (LHC) features,
with flux densities ranging from 0.7 to 7.1 Jy beam$^{-1}$. The LSR
velocities of these features range from 13.8 to 15.4 km s$^{-1}$.

This paper represents the first high resolution observations of
6.668-GHz methanol maser emission in ON1 combined with the first
detailed studies of the polarization properties of both the 6.668-GHz
Methanol and 6.031/6.035-GHz excited-state OH in ON1. Unfortunately,
the MERLIN results presented here did not have a wide enough bandwidth
both to give adequate spectral resolution and to cover both features,
so they were centred near the main peak shown in
Fig. \ref{figure1}. 

\section[]{OBSERVATIONS}
ON1 was observed using the Multi Element Radio Linked Interferometer
Network (MERLIN) on 2005 January 12 and 16. These were the first data
taken with new broadband 4-8 GHz e-MERLIN receivers on 5 telescopes
(the MKII, Darnhall, Tabley, Knockin and Cambridge), enough for full
synthesis imaging, in full polarization.  Observations were taken of
three maser transitions: 6030.747 MHz (F = 2 $-$ 2 hyperfine
transition of the $^{2}$$\Pi$$_{3/2}$, J = 5/2 excited-state OH);
6035.092 MHz (F = 3 $-$ 3 hyperfine transition of the
$^{2}$$\Pi$$_{3/2}$, J = 5/2 excited-state OH); and 6668.5192 MHz
(5$_{1}$$\rightarrow$6$_{0}$ transition of A$^{+}$ methanol).  The
longest baseline of MERLIN is 217 km, giving a synthesized beam size
of 47 mas at 6.0 GHz and 43 mas at 6.7 GHz.  Spectral line data were
taken using a 0.5 MHz bandwidth, centred on each line frequency
corrected to a V$_{\rm LSR}$ of 12 km s$^{-1}$, and split into 512
frequency channels. This gave a velocity span and resolution of 24.9
km s$^{-1}$ in 0.049 km s$^{-1}$ channels at 6.0 GHz and 22.5 km
s$^{-1}$ in 0.044 km s$^{-1}$ channels at 6.7 GHz. To obtain all the
Stokes parameters, both right and left hand circularly polarized
signals were collected from each telescope in the network.  The phase
reference source 2013+340 ($\approx$3$^{\circ}$ from ON1) was observed
in continuum mode with 15 MHz useable bandwidth, for 2 min, before and
after 7-min scans on ON1 at each frequency. Each line plus reference
was observed in rotation, for a total of 5 hrs on ON1 at each OH
frequency and 3.5 hrs at the methanol frequency.  3C286 and 3C84 were
observed at all frequencies in both wide- and narrow-band
configurations for totals of 6.0 and 9.2 hrs, respectively (with an
approximately even split in time between the configurations).

Using local MERLIN software, the flux density of 3C84 was established
to be 15.5 Jy with respect to the primary amplitude calibration
source, 3C286.  The data were converted to FITS files, using 3C84 to
apply a preliminary bandpass scaling to all data.  All further
processing was performed using the Astronomical Image Processing
Software ({\sc aips}), see the MERLIN User Guide (Diamond et al. 2003)
for details. A small percentage of the data had anomalous phase and
amplitude and was edited accordingly.  Phase self-calibration was
performed for all the calibration sources and 3C84 was used to derive
the instrumental phase offset between the continuum and spectral
configurations.  Amplitude self-calibration was performed on 3C84 and
this source was used to derive bandpass correction tables and also to
measure the polarization leakage. Systematic leakage was calibrated
and there was found to be a residual leakage of $\le$ 0.5 per cent
Stokes I in Stokes Q and U and $\le$ 0.25 per cent in Stokes V.  3C286
was used to derive the polarization angle correction.  The various
corrections appropriate for each frequency were then applied to ON1,
including phase and amplitude solutions derived from 2013+340.

Inspection of the spectra of each of the ON1 lines in LHC and RHC
enabled the selection of the brightest channel in a single hand of
circular polarization, this was imaged to obtain an accurate reference
position.  The systematic position errors are $\sim$12 mas at 6$-$7
GHz due to uncertainty in the phase calibration, telescope positions
and atmospheric variation (Etoka, Cohen \& Gray 2005).  The clean
components of each reference image were then used as a model for phase
and amplitude self-calibration and the solutions applied to all
channels and both polarizations of that line. Finally, the brightest
channel in the weaker hand of polarization for each line was selected
and self-calibrated, applying the solutions to that hand of
polarization only. In this way, good calibration was achieved without
compromising the absolute position accuracy or the polarized phases
and amplitudes.

When all calibration was complete, the contributions of the antennas
were re-weighted according to their sensitivity, their individual
efficiency and to avoid baselines to a single antenna completely
dominating the array (http://www.merlin.ac.uk/user\_guide/).  This
allowed preparation of clean image datacubes for each line in all 4
Stokes parameters (I, Q, U and V) and in RHC and LHC. A circular
restoring beam of 50 mas FWHM was used with a pixel size of 12 mas,
giving in total an image field of view of 12 arcsec. The Stokes I
images had a 1$\sigma_{\rm rms}$ noise level of 25 mJy beam$^{-1}$ and
15 mJy beam$^{-1}$ for methanol and OH respectively, except in the
brightest total intensity channels, which were dynamic range limited
with a noise level of up to 0.5 per cent of the peak.  The LHC and RHC
images had noise levels of 35, 20 and 25 mJy beam$^{-1}$ for the
methanol, 6.031- and 6.035-GHz OH lines, respectively.

Elliptical Gaussian components were fitted to each patch of emission
above 5$\sigma_{\rm rms}$ in each Stokes I, LHC and RHC channel and
the position, deconvolved size, peak and total intensities, and the
uncertainties were all measured.  Any components not found at similar
positions in a series of 5 or more consecutive channels were
discarded.  The relative positional errors for all components,
determined from the errors in the peak position of the Gaussian fits,
were 2 mas or better (for the accuracy of Gaussian component fitting
see Condon 1997; Condon et al. 1998; Richards, Yates \& Cohen 1999).
The uncertainty in component size is the square root of 2 multiplied
by the position uncertainty, which enables brighter component sizes to
be determined with meaningful accuracy.  The position and intensity of
emission in the Q, U, and V channel maps were measured at the position
of each Stokes I component for that line and channel.

\begin{figure*}
\includegraphics[width=18cm]{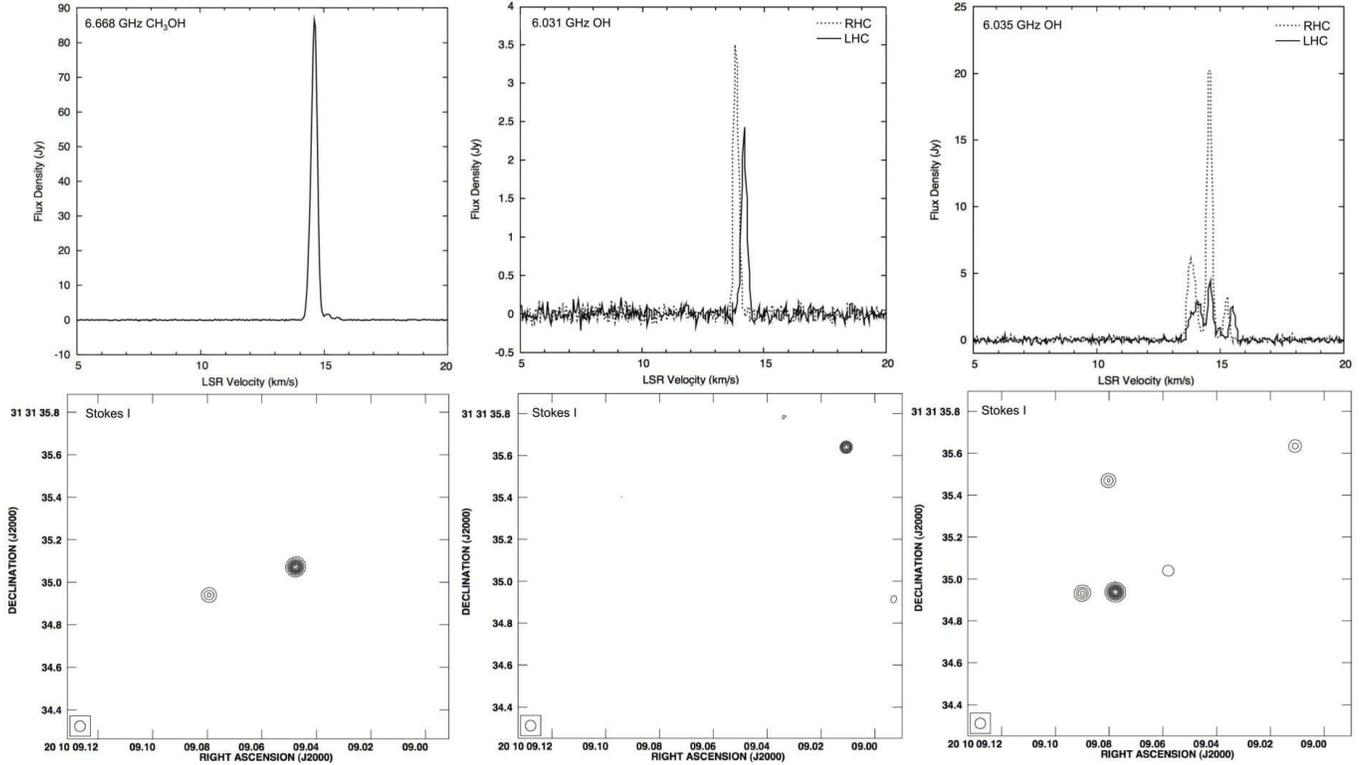}
\caption{\small From left to right: Stokes I 6.668-GHz methanol
emission and map, RHC and LHC (dotted and solid lines respectively)
6.031-GHz OH emission and Stokes I map, RHC and LHC 6.035-GHz OH
emission and Stokes I map. Maps show emission integrated over all
channels, with contour levels at 10 to 100 per cent of the peak flux
density for the 6.667-GHz methanol and 6.035-GHz OH emission and 40 to
100 per cent for the 6.031-GHz OH emission.  The peaks are 72.76 Jy
beam$^{-1}$, 7.87 Jy beam$^{-1}$ and 1.71 Jy beam$^{-1}$
respectively.}
\label{figure4}
\end{figure*}

\section[]{RESULTS}
\subsection{6.7-GHz Methanol}
51 individual channel Stokes I components were detected in the
6.667-GHz methanol transition.  These individual channel components
were grouped into features as described in section 2 and the flux
density weighted mean properties for each feature are given in Tables
\ref{table1} and \ref{table1.5}.  Four methanol features were observed
with peak velocities between 14.46 km s$^{-1}$ and 15.57 km s$^{-1}$.
The positions and velocities of the 6.7-GHz methanol features are
given in Table \ref{table1} and the Stokes I spectrum and
channel-averaged map are shown in Fig. \ref{figure4}.  Features show
typical FWHM linewidths of between 0.21 and 0.28 km s$^{-1}$. Peak
flux densities of the individual features varied between 0.48 Jy
beam$^{-1}$ and 73.62 Jy beam$^{-1}$.  Estimates of the deconvolved
component sizes from the Gaussian fitting ranged from 16 to 52 mas
(equivalent to 29 to 94 AU at the assumed distance of 1.8
kpc). Combined with the peak flux densities these allowed lower limits
to the peak brightness temperatures to be determined (ranging between
1.99 $\times$ 10$^{7}$ K and 8.97 $\times$ 10$^{9}$ K).

No previous high resolution observations of 6.7-GHz methanol are
available in the literature for comparison.

A larger restoring beam of 100 mas was used to search for extended
emission, similar to that seen in W3(OH) by Harvey-Smith \& Cohen
(2005), but none was found. It is difficult to speculate how much of
the single dish flux density is recovered by the current observations
as the previous single-dish observations were taken in 1999 (Szymczak
et al. 2000) and showed possible variability in comparison to the
previous observations in 1991 (Menten 1991). The two flux densities
recorded were, as noted in Section 1, 109 Jy and 91 Jy respectively,
taken with the 32-m Torun and the 43-m Greenbank telescope. The errors
in the absolute flux density calibrations for the two observations
were $\pm$15 per cent and $\pm$10 per cent respectively, so the two
measurements could be consistent to within the errors. However, there
have been a number of studies into the variability of 6.7-GHz methanol
masers, most recently by Goedhart, Gaylard \& van der Walt (2005), and
these have served to show that 6.7-GHz methanol masers are generally
variable on timescales of years, but the nature of their variability
is wide ranging from periodic/quasi-periodic through monotonic
increases/decreases to sporadic behaviour, and so with just two
previous epochs it is impossible to judge the nature of the possible
increased flux density in 1999. Given there is a lack of extended
emission from the 100 mas restoring beam and if minimal variability is
assumed, then comparison with the spectrum of Szymczak et al. (2000)
implies the flux density recovered is likely to be over 80 per cent.
 
\subsection{6-GHz Excited-state OH}
14 individual channel Stokes I components were detected in the
6.031-GHz OH transition, and 63 for that at 6.035-GHz OH. 9 individual
channel LHC and 8 RHC components were detected for 6.031-GHz OH,
whilst 53 individual channel LHC components and 50 RHC, were found for
6.035-GHz OH.  These individual channel components were grouped into
features as described in section 2 and the flux density weighted mean
properties for each feature are given in Tables \ref{table2} and
\ref{table3}.  A single OH LHC feature and a single OH RHC feature
were found at 6.031 GHz with peak velocities of 14.19 km s$^{-1}$ and
13.87 km s$^{-1}$ respectively. 6 OH LHC and 5 OH RHC features were
found at 6.035 GHz with peak velocities between 13.76 km s$^{-1}$ and
15.50 km s$^{-1}$. The positions and velocities of these OH features
are given in Table \ref{table2}, with the corresponding Stokes I
feature details in Table \ref{table3}. The RHC and LHC spectra and
Stokes I channel-averaged maps are shown in
Fig. \ref{figure4}. Linewidths vary between 0.15 and 0.34 km
s$^{-1}$. The peak flux densities varied between 0.35 Jy beam$^{-1}$
and 15.72 Jy beam$^{-1}$.  Estimates of the deconvolved component
sizes from the Gaussian fitting ranged from 13 to 38 mas (equivalent
to 23 to 68 AU at the assumed distance of 1.8 kpc). Combined with the
peak flux densities these allowed lower limits to the peak brightness
temperatures to be determined (ranging between 1.02 $\times$ 10$^{7}$
K and 2.71 $\times$ 10$^{9}$ K).

The overall shape of the 6.035-GHz OH emission spectrum is consistent
with the VLBI spectrum of Desmurs \& Baudry (1998) and the single-dish
spectrum of Baudry et al. (1997), with peaks at similar velocities and
similar relative intensities between the spectral features. However
the overall flux density in the current observations is
greater. Baudry et al. (1997) established from two epochs, separated
by a year, that there is rapid time variability of the source, and
hence over the 10 years between those observations and the current,
significant flux density variation could have occurred and, similar to
the methanol, any derivation of the recovered flux density would lack
significant accuracy.

\begin{figure*}
 \centering
\includegraphics[width=17.5cm]{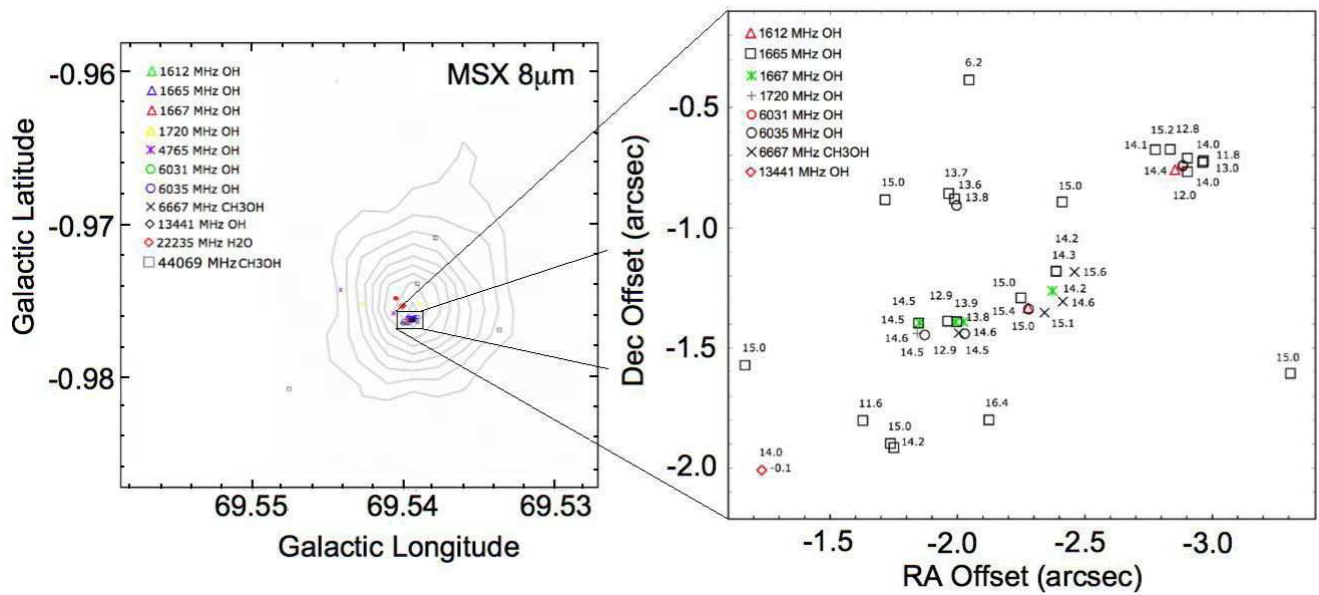}
\caption{\small Left: MSX 8 micron emission from the online data
archive (http://irsa.ipac.caltech.edu/data/MSX/) with a peak flux
density of 0.6 $\times$ 10$^{-5}$ W m$^{-2}$ sr$^{-1}$, with overlayed
positions of known masers in the region. There are: Ground-state OH
from Nammahachak et al. (2006), 4.765-GHz OH from Baudry \& Diamond
(1991), 13.4-GHz OH from Baudry \& Desmurs (2002), H$_{2}$O maser
positions from Downes et al. (1979) and 44-GHz methanol from Kurtz et
al. (2004); Right: The new excited-state OH and methanol features
together with the known ground-state positions of Nammahachak et
al. (2006) and the 13.4-GHz OH of Baudry \& Desmurs (2002). Offset
positions on the right are given relative to the MSX point source at
RA (J2000) = 20$^{h}$ 10$^{m}$ 09$^{s}$.23628, Dec (J2000) =
31$^{\circ}$ 31$'$ 36$''$.3784. LSR velocities are listed next to the
maser positions.}
\label{figure2}
\end{figure*}

Comparison between the features found by Desmurs \& Baudry (1998) and
the current observations is also difficult, again due to the
variability, but also due to the error in absolute position of the
previous observations (between 0.2 and 0.5 arcsec). However the range
of velocities of the features between 13.6 and 15.4 km s$^{-1}$
is similar, as are the relative positions of the 6.035- and 6.031-GHz OH
emission (described in Section 4 and seen in Fig. \ref{figure2}).
Specifically at 6.031-GHz OH, Fish et al. (2006) and Desmurs \& Baudry
(1998) found 1 LHC and 1 RHC feature at velocities of 14.2 and 13.8 km
s$^{-1}$ respectively, which are also found in the current
observations. For the 6.035-GHz OH, the current observations fail to
detect the LHC features at 5.47, 7.74 and 12.59 km s$^{-1}$ seen by
Fish et al., but do detect 2 previously unseen LHC features.  The
current study also fails to detect the RHC feature at 5.53 km
s$^{-1}$, but otherwise recovers the RHC features seen by Fish et
al. and Desmurs \& Baudry and finds 1 further feature.

\subsection{Linear Polarization}
\begin{figure}
 \centering
\includegraphics[width=8.5cm]{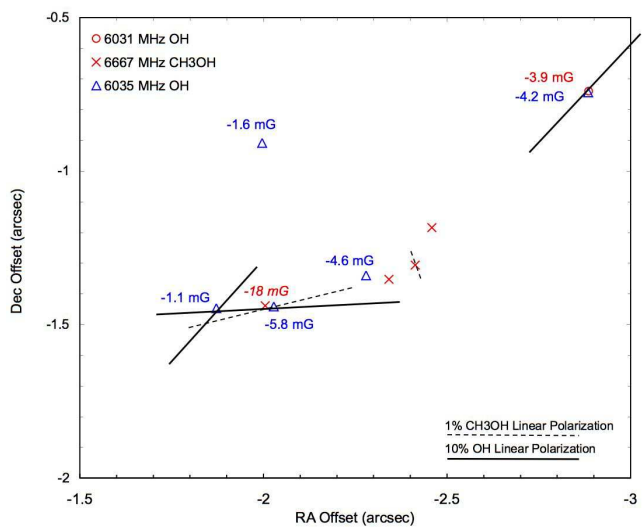}
\caption{\small Linear polarization properties for the statistically
  significant 6.031-GHz OH and 6.035-GHz OH and the upper limits of
  the 6.668-GHz methanol features. Numbers indicate the magnetic field
  strengths deduced from Zeeman Splitting, with blue representing
  6.035-GHz OH, the $-$3.9 mG in red is the 6.031-GHz field and the
  $-$18 mG in red italics is the tentative 6.667-GHz methanol
  detection. Positions are given relative to RA (J2000) = 20$^{h}$
  10$^{m}$ 09$^{s}$.23628, Dec (J2000) = 31$^{\circ}$ 31$'$
  36$''$.3784.}
\label{figure3}
\end{figure}

Two of the four 6.7-GHz methanol features demonstrate, in terms of
random noise, statistically significant ($>$3$\sigma$) Stokes Q and U
flux densities, resulting in linear polarizations of 0.2 per cent and
1.3 per cent. However, analysis of the bandpass calibrator 3C84,
showed a residual polarization leakage of $\le$0.5 per cent, reducing
the significance of the measurements to an upper limit.  The 6.031-GHz
OH Stokes I feature had significant Q and U flux density giving 12 per
cent linear polarization. Of the five 6.035-GHz OH Stokes I features,
two had significant Q and U flux density giving linear polarizations
of 19 per cent and 10 per cent. The 3 features that had statistically
significant linear polarization showed reasonably consistent
polarization angles of $-$60$^{\circ}$ $\pm$28$^{\circ}$.  The upper
limit of $\sim$1 per cent linear polarization in methanol is
consistent with that found in W3(OH) by Vlemmings et al. (2006a) and
in NGC6334F by \cite{Elling02}, where both had the majority of features
displaying less than 5 per cent linear polarization.  Linear
polarization of excited-state OH in ON1 has not been studied before.
The polarization vectors of the statistically significant OH results
are plotted in Fig. \ref{figure3} together with the upper limits of
methanol polarization.

The two methanol features with tentative linear polarization have a
90$^{\circ}$ difference in polarization angle.  As shown in the case
of SiO and H$_{2}$O masers, which are also diamagnetic, such a
90$^{\circ}$ flip can be caused by a difference in the angle $\theta$
between the maser line of sight and the magnetic field. When $\theta$
is larger than the critical angle $\theta_{\rm
crit}\approx55^{\circ}$, the linear polarization direction is
perpendicular to the magnetic field, whilst when $\theta<\theta_{\rm
crit}$, the linear polarization is parallel. Furthermore, as the
fractional linear polarization also decreases close to the critical
angle, this could explain why the strongest methanol maser feature has
the lowest significant linear polarization fraction (Vlemmings et
al. 2006b, Vlemmings \& Diamond 2006, references therein).

\subsection{Circular Polarization \& Magnetic Fields}
In order to investigate the degree of circular polarization within
methanol the cross-correlation method of Modjaz et al. (2005) was
employed. This method estimates Zeeman splitting via the
cross-correlation of the RR and LL spectra and as such is independent
of any polarization leakage that may affect the Stokes V spectrum.
Also, assuming the magnetic field strength is similar across the
blended maser features, it allows for a determination of Zeeman
splitting when spectral blending makes a measurement using the V
spectrum impossible. The rms of the cross-correlation method is a
function of the spectral channel width and the rms noise on the RR and
LL spectra. The method was shown by Modjaz et al. (2005), via Monte
Carlo simulations, to have a sensitivity equivalent to the standard
S-curve method. Through this method, methanol feature D is seen to
have a Zeeman splitting of 0.0009$\pm$0.0003 km s$^{-1}$. Compared
with OH, methanol is believed to have a much lower Zeeman splitting
coefficient, but a precise value has never clearly been defined. Using
the only currently available value of the methanol g-Land\'e factor,
determined by Jen (1951) from 25 GHz methanol maser lines, the Zeeman
splitting coefficient of the 6.7-GHz methanol maser transition is
calculated to be 0.0493 km s$^{-1}$ G$^{-1}$ (Vlemmings et
al. 2006a). Consequently the Zeeman splitting of feature D corresponds
to a field strength of $-$18$\pm$6 mG. This represents the first
tentative detection of Zeeman splitting in 6.7-GHz methanol. As the
noise is limited by the dynamic range, the noise increases in the peak
channels to $\sim$60 mJy, and therefore means the field strength is at
$\sim$3$\sigma$ (rather than $>$5$\sigma$, as it would have been
without the dynamic range limitation). In terms of fractional circular
polarization, statistically significant results could not be
determined due to the residual leakage, but the feature demonstrating
the Zeeman splitting had a circular polarization of 0.65 per cent at
2.5 $\sigma$. 

In total six excited-state OH Zeeman pairs were found, five at 6.035 GHz
and one at 6.031 GHz, with the LHC and RHC features in each pair having
a spatial association to within 1 mas. Splitting factors of 0.0564 km
s$^{-1}$ mG$^{-1}$ and 0.0790 km s$^{-1}$ mG$^{-1}$, respectively,
were assumed (Yen et al. 1969). This implies a magnetic field strength
ranging from $-$1.1 to $-$5.8 mG for the pairs at 6.035 GHz and $-$3.9
mG for the single pair at 6.031 GHz.  All the magnetic fields are
directed towards us and concur in magnitude and direction with those
found previously by \cite{Desmurs98}, which varied between $-$3.6 mG
and $-$6.3 mG, and the single-dish measurements of Fish et al. (2006)
of between $-$0.8 mG and $-$5.0 mG. Furthermore the fields seen in ON1
appear to be typical for excited-state OH in star-formation regions in
general, which have been observed to vary between $+$9.1 mG and
$-$13.5 mG (Fish et al. 2006).  Fig. \ref{figure3} shows the location
of the measured field strengths.  Overall the fields compare
favourably with the field strengths found for the ground-state OH by
\cite{Namma06}, which varied between $-$0.4 and $-$4.6 mG.

\begin{table*}
\begin{minipage}{160mm}
\small
\centering
\caption{\small Properties for the 6.7-GHz methanol features in the
ON1 star-forming region. Positions relative to RA (J2000) = 20$^{h}$
10$^{m}$ 09$^{s}$.23628, Dec (J2000) = 31$^{\circ}$ 31$'$
36$''$.3784. Features A and B have relative positions accurate to
$\pm$0.001 arcsec or better, whilst C and D have an accuracy of
$\pm$0.002 arcsec.  As the velocities are taken as the mid-point of
channels there is an error of $\pm$0.5 channels (corresponding to
$\sim$0.02 km s$^{-1}$).}
\begin{tabular}{c r r r r c}
\\
\hline
Feat. & \multicolumn{1}{c}{$\Delta$RA} &  \multicolumn{1}{c}{$\Delta$Dec}
& \multicolumn{1}{c}{V$_{\rm LSR}$} & \multicolumn{1}{c}{$\Delta$V$_{1/2}$} 
& \multicolumn{1}{c}{Peak T$_{\rm b}$}
\\
 & (arcsec) & (arcsec)  & (km/s) & (km/s) & (K)
 \\
\hline 
A & $-$2.459 & $-$1.184 & 15.57 & 0.21 & $\ge$1.99 $\times$ 10$^{7}$ \\
B & $-$2.342 & $-$1.353 & 15.13 & 0.28 &  $\ge$1.83 $\times$ 10$^{8}$ \\
C & $-$2.413 & $-$1.307 & 14.62 & 0.27 & $\ge$8.97 $\times$ 10$^{9}$ \\
D & $-$2.005 & $-$1.438 & 14.46 & 0.27 & $\ge$3.43 $\times$ 10$^{9}$ \\
\end{tabular} 
\label{table1}
\end{minipage}
\end{table*}

\begin{table*}
\begin{minipage}{160mm}
\small
\centering
\caption{\small Stokes parameters flux densities and polarization
properties for the 6.7-GHz methanol features in the ON1 star-forming
region.  Stokes I, Q, U and V have a noise rms of $\sim$25
mJy. Polarization intensity is accurate to approximately $\pm$5 per
cent. In addition to the listed polarization angle errors there is an
additional possible systematic error of 5$^{\circ}$ due to
calibration.  Figures in italics represent less than 3$\sigma$$_{\rm
rms}$ significance.}
\begin{tabular}{c r r r r r r }
\\
\hline
Feat. & \multicolumn{1}{c}{I} & \multicolumn{1}{c}{Q} & \multicolumn{1}{c}{U} & \multicolumn{1}{c}{V} & \multicolumn{1}{c}{P}
& \multicolumn{1}{c}{$\chi$} 
\\
 & (Jy/b) & (Jy/b)& (Jy/b)& (Jy/b)& (Jy/b)&\multicolumn{1}{c}{($^{\circ}$)}
 \\
\hline 
A & 0.46 & \textit{0.00} & \textit{0.00} & \textit{0.01} & \textit{0.02} & $-$ \\
B & 1.08 & \textit{0.00} & \textit{$-$0.01} & \textit{0.00} & \textit{0.03} & $-$ \\
C & 53.56 & 0.09 & 0.08 & 0.14 & 0.13 & 20.6$\pm$2.0 \\
D & 20.03 & $-$0.23 & $-$0.10 & 0.13 & 0.26 & $-$76.7$\pm$2.0 \\
\end{tabular} 
\label{table1.5}
\end{minipage}
\end{table*}

\begin{table*}
\begin{minipage}{160mm}
\small
\centering
\caption{\small Hydroxyl maser features for the ON1 star-forming
region. Positions relative to RA (J2000) = 20$^{h}$ 10$^{m}$
09$^{s}$.23628, Dec (J2000) = 31$^{\circ}$ 31$'$ 36$''$.3784.  All
relative positions are accurate to $\pm$0.001 arcsec or better, with
the exception of feature B which is $\pm$0.002 arcsec.  Velocities are
accurate to $\pm$0.02 km s$^{-1}$.  Peak flux density is accurate to
$\sim$0.02 Jy beam$^{-1}$. The magnetic field strengths are accurate
to 0.4 mG.}
\begin{tabular}{l c c c c c c c c l}
\\
\hline
Transition & \multicolumn{1}{c}{Feat.} & \multicolumn{1}{c}{Stokes}  & \multicolumn{1}{c}{$\Delta$RA} &  \multicolumn{1}{c}{$\Delta$Dec}
& \multicolumn{1}{c}{V$_{\rm LSR}$} & \multicolumn{1}{c}{$\Delta$V$_{1/2}$} & \multicolumn{1}{c}{Peak Flux Density} & \multicolumn{1}{c}{Peak T$_{\rm b}$}
& \multicolumn{1}{l}{Comment}
\\
 &  & I No. & (arcsec) & (arcsec) & (km/s) & (km/s) & (Jy/b) & (K)
 \\
\hline 
6031-LHC & A & 1 & $-$2.886  & $-$0.739  & 14.19 & 0.27 & 2.34 & $\ge$9.45 $\times$ 10$^{8}$ & Z$_{1}$ $-$3.9mG\\ 
6031-RHC & a & 1 & $-$2.885  & $-$0.739  & 13.87 & 0.25 & 3.46 & $\ge$1.29 $\times$ 10$^{9}$ & Z$_{1}$\\
6035-LHC & A & 1  & $-$2.280  & $-$1.339  & 15.50 & 0.24 & 2.32 & $\ge$3.83 $\times$ 10$^{8}$ & Z$_{2}$ $-$4.6mG\\
& B & - & $-$2.234  & $-$1.352  & 15.06 & 0.15 & 0.35 & $\ge$1.02 $\times$ 10$^{7}$ & \\
& C & 2 & $-$2.028  & $-$1.440  & 14.89 & 0.29 & 0.51 & $\ge$3.04 $\times$ 10$^{7}$ & Z$_{3}$ $-$5.8mG\\
& D & 3 & $-$1.871  & $-$1.446  & 14.57 & 0.23 & 3.71 & $\ge$4.58 $\times$ 10$^{8}$ & Z$_{4}$ $-$1.1mG\\
& E & 4 & $-$2.884  & $-$0.744  & 14.16 & 0.34 & 2.23 & $\ge$4.56 $\times$ 10$^{8}$ & Z$_{5}$ $-$4.2mG\\
& F & 5 & $-$1.996  & $-$0.907  & 13.85 & 0.32 & 1.89 & $\ge$1.67 $\times$ 10$^{8}$ & Z$_{6}$ $-$1.6mG\\
6035-RHC & a & 1 & $-$2.280  & $-$1.339  & 15.25 & 0.26 & 2.99 & $\ge$4.97 $\times$ 10$^{8}$ & Z$_{2}$\\
 & b & 2 & $-$2.028  & $-$1.441  & 14.56 & 0.23 & 15.72 & $\ge$2.71 $\times$ 10$^{9}$ & Z$_{3}$\\
 & c & 3 & $-$1.872  & $-$1.446  & 14.51 & 0.22 & 3.86 & $\ge$4.53 $\times$ 10$^{8}$ & Z$_{4}$\\
 & d & 4 & $-$2.884  & $-$0.744  & 13.92 & 0.30 & 3.34 & $\ge$1.12 $\times$ 10$^{9}$ & Z$_{5}$\\
 & e & 5 & $-$1.996  & $-$0.908  & 13.76 & 0.24 & 3.66 & $\ge$3.64 $\times$ 10$^{8}$ & Z$_{6}$\\
\end{tabular} 
\label{table2}
\end{minipage}
\end{table*}

\begin{table*}
\begin{minipage}{180mm}
\small
\centering
\caption{\small Stokes parameters flux densities and polarization
properties for the excited-state OH features in the ON1 star-forming
region. Positions relative to RA (J2000) = 20$^{h}$ 10$^{m}$
09$^{s}$.23628, Dec (J2000) = 31$^{\circ}$ 31$'$ 36$''$.3784. Both the
6.031-GHz features and 6.035-GHz features have relative positions
accurate to $\pm$0.001 arcsec.  Stokes I, Q, U and V have an rms noise
of $\sim$15 mJy for both transitions. Polarization intensity is
accurate to approximately $\pm$5 per cent.  In addition to the listed
polarization angle errors there is an additional possible systematic
error of 5$^{\circ}$ due to calibration.  As the velocities are taken
as the mid-point of channels there is an error of $\pm$0.5 channels
(corresponding to $\sim$0.02 km s$^{-1}$). Figures in italics
represent less than 3$\sigma$ significance.}
\begin{tabular}{l r r r r r r r r r r r r}
\\
\hline
No. & \multicolumn{1}{c}{$\Delta$RA} &  \multicolumn{1}{c}{$\Delta$Dec}
& \multicolumn{1}{c}{V$_{\rm LSR}$} 
& \multicolumn{1}{c}{I} & \multicolumn{1}{c}{Q} & \multicolumn{1}{c}{U} & \multicolumn{1}{c}{V} & \multicolumn{1}{c}{P}
& \multicolumn{1}{c}{$\chi$} & \multicolumn{1}{c}{m$_{\rm l}$} & \multicolumn{1}{c}{m$_{\rm c}$} & \multicolumn{1}{c}{m$_{\rm t}$}
\\
 & (arcsec) & (arcsec)  & (km/s) & (Jy/b)
 & (Jy/b)& (Jy/b)& (Jy/b)& (Jy/b)&\multicolumn{1}{c}{($^{\circ}$)}&(per cent)&(per cent)&(per cent)
 \\
\hline 
\multicolumn{2}{l}{6.031 GHz}\\
1 & $-$2.885 & $-$0.739 & 13.99 & 1.25 & \textit{0.01} & $-$0.15 & 0.41 & 0.20 & $-$43.1$\pm$1.9 & 12.2$\pm$2.4 & 33.0$\pm$2.7 & 35.2$\pm$3.3\\
\multicolumn{2}{l}{6.035 GHz}\\
1 & $-$2.279 & $-$1.339 & 15.35 & 1.29 & \textit{$-$0.01} & \textit{$-$0.01} & 0.22 & \textit{0.03} & $-$ & $-$ & $-$ & $-$\\
2 & $-$2.028 & $-$1.441 & 14.57 & 6.29 & $-$1.16 & $-$0.07 & 5.30 & 1.23 & $-$87.7$\pm$2.0 & 18.5$\pm$0.5 & 84.3$\pm$0.7 & 86.3$\pm$0.8\\
3 & $-$1.871 & $-$1.446 & 14.57 & 2.89 & \textit{0.02} & $-$0.29 & 0.10 & 0.29 & $-$42.5$\pm$0.7 & 10.0$\pm$1.0 & 3.4$\pm$0.9 & 10.6$\pm$1.2\\
4 & $-$2.884 & $-$0.744 & 14.02 & 1.62 & \textit{$-$0.03} & \textit{$-$0.02} & 0.42 & 0.09 & $-$ & $-$ & $-$ & $-$\\
5 & $-$1.996 & $-$0.907 & 13.80 & 1.93 & \textit{0.01} & \textit{$-$0.01} & 0.52 & \textit{0.03} & $-$ & $-$ & $-$ & $-$\\
\end{tabular} 
\label{table3}
\end{minipage}
\end{table*}


\section[]{DISCUSSION}
\subsection{Distribution \& Coincidences}
The methanol features lie in a roughly linear south-east to north-west
distribution, covering about 0.6 arcsec ($\sim$1080 AU at 1.8 kpc).
The excited-state OH features meanwhile show a wider spread of about
1.2 arcsec ($\sim$2160 AU), with one feature offset from the linear distribution by
about 0.5 arcsec ($\sim$900 AU). The linear distribution of both the
6.7-GHz methanol and the 6-GHz excited-state OH is parallel to the
mainline ground-state OH distribution of Nammahachak et al. (2006),
with the methanol consistently offset by on average $\sim$70 mas (130
AU). The three 6.035-GHz features towards the centre of the
distribution, seen in Fig. \ref{figure2}, are systematically offset
from the ground-state 1665-MHz OH features by an average of $\sim$57
mas (100 AU).

It is worth considering if proper motion of the maser features between
the observations of Nammahachak et al.  observed in 1996 and the
current observations in 2005 could explain the observed spatial offset
between the ground-state OH and the methanol and excited-state OH
maser features (both sets of observations used the same extragalactic phase
reference source). The systematic offset between the 6.035-GHz OH and 1665-MHz
OH of 100 AU would require a motion over $\sim$1.5 $\times$ 10$^{10}$
km in 9 years, which is $\approx$50 km s$^{-1}$. Both internal and
external proper motion should be examined.  Work by Bloemhof, Reid \&
Moran in 1992 looked at the internal proper motion of the 1665-MHz OH
masers in the W3(OH) region, which is located at a similar distance,
and found very few motions greater than 5 mas over the 7.5 year period
they studied (velocities were typically a few km s$^{-1}$). This
implies that internal proper motion is unlikely to account for the
separation seen.  It is also likely that any internal proper motion of
the ground-state OH masers would also affect the excited-state masers,
as their masing gas clouds are likely to share a bulk proper motion.
In terms of external proper motion it is possible to examine the
effect of Galactic rotation on the region over the 9 year
separation. Adopting the rotation curve of Brand \& Blitz (1993),
leads to an external east to west motion of $\sim$45 km s$^{-1}$ for
the region, which would amount to $\sim$45 mas over 9 years. This
would result in a separation of $\sim$ 10 mas between the overall line
of ground-state and excited-state OH masers and $\sim$ 25 mas between
the ground-state OH and the methanol. Coupled with the uncertainties
in these calculations, possible peculiar motion, and the absolute
positional errors of the two sets of data (20 mas and 15 mas
respectively), it is possible that proper motion may account for the
spatial separation seen. An accurate determination of these offsets
requires near simultaneous observations of both frequency regimes, but
this is beyond the scope of the current paper.

\begin{figure}
 \centering
\includegraphics[width=9cm]{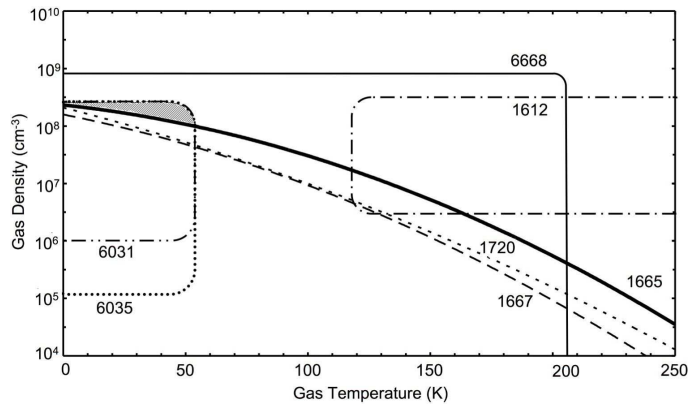}
\caption{\small Approximate parameter space plot of gas kinetic
temperature versus gas density for the OH and methanol maser
transitions. Allowed regions are those enclosed by the lines and the
axes. The shaded region represents the temperatures and densities for
which excited-state OH masers are seen, but ground-state OH is not.
The plot is adapted from the results of Cragg, Sobolev \& Godfrey
(2002) and Cragg, Sobolev \& Godfrey (2005), based on a dust
temperature of 175 K. It must be noted that several other factors not
incorporated in this plot are important for maser emission, such as
velocity-coherent column density, fractional number density and
optical depths for the IR radiation.}
\label{figurePP}
\end{figure}

However, if this apparent lack of spatial coincidence is demonstrated
it could be the result of differences in the specific column densities
of the two species, as modeling of OH and methanol masers by Cragg,
Sobolev \& Godfrey (2002) showed that both OH and methanol require
very similar high density, low temperature regimes (dust temperatures
exceeding 100 K, gas temperatures $<$100 K, and densities in the range
10$^{5}$ $<$ n$_{\rm H}$ $<$ 10$^{8.3}$ cm$^{-3}$). On the other hand,
subtle variations in the above parameters could pick out the exact
conditions required for each species of maser. A plot of the
approximate regions for maser emission for a dust temperature of 175 K
is given in Fig. \ref{figurePP}.

For the excited-state OH and methanol frequencies of the current
observations, based on a combination of the absolute and relative
positional errors as per Etoka et al. (2005), 15 mas can be taken as
the distance for association, within which it is not possible to
determine if individual methanol and excited-state OH features are
spatially separate and hence may indicate coincidence and/or
co-propagation.  In the current study, there is no association between
the 6.7-GHz methanol and the 6-GHz excited-state OH (the closest
emission peaks are separated by $\sim$23 mas, even taking into
account the deconvolved component sizes there is still distinct
separation). The two transitions also display significantly different
magnetic field strengths as described in section 4.3.
 The RHC and LHC Zeeman pair at 6.031 GHz coincides with
a 6.035-GHz OH Zeeman pair to $<$5.5 mas spatially and $<$0.049 km
s$^{-1}$ in velocity, implying co-propagation. This is a confirmation
of the suspected coincidence which was identified in Desmurs \& Baudry
(1998) to within their accuracy of $\sim$200 mas.  Gray, Field \& Doel
(1992) show that for the two transitions to be coincident, if the dust
temperature is $\sim$50 K and assuming a hydrogen number density of
2.5 $\times$ 10$^{7}$ cm$^{-3}$, then the kinetic temperature must be
$\sim$75 K. However the results of Cragg et al. (2002), using a more
extensive model and a dust temperature of 175 K, allow for a range of
kinetic temperatures (as shown in Fig. \ref{figurePP}).  The 100 per
cent association of 6.031-GHz emission with 6.035-GHz is consistent
with previous results, such as by Etoka et al. (2005).  The fact that
there is only one case of coincidence of all the features shows the
physical conditions of the main group of maser spots must vary from
the offset coincident pair.

At face value the data sets we have show that the individual 6-GHz
excited-state OH features do not appear to show any spatial
coincidence with ground-state OH to within 20 mas (the closest
separation of peak emission is $\sim$28 mas). There is one exception,
a satellite line feature at 1612-MHz.  There are only two 1612-MHz OH
features in the region, one is separated by 34 mas from 6.035-GHz
emission, the other has just a 7 mas separation. As mentioned earlier
this may be accounted for by proper motion of the region, but if not
then it may imply, contrary to previous models, that there might be an
overlap in the conditions for maser emission between the 1612-MHz
transition and the 6.035-GHz transition, which is not present for the
other ground-state OH transitions.  On the other hand the two
transitions could be tracing higher density gas, but not be spatially
coincident, and so exist in different temperature regions.  Gray et
al. (1992) suggest the 1612-MHz transition requires kinetic gas
temperatures of $\geq$150 K and hydrogen number densities around 6
$\times$ 10$^{6}$ cm$^{-3}$. Fig. \ref{figurePP} also implies a higher
density is required for both to be present, but if the dust
temperature is consistent, the kinetic temperature regimes vary (but
as noted in the caption to the Figure other factors may allow for the
two to be coincident).

It is also possible to conclude that the possible spatial separation
between the ground-state and excited-state OH (with the exception of
the one 1612-MHz feature), could mean we see the very highest density
of OH in the excited-state regions, as the conditions for
excited-state OH maser emission extend to higher densities than the
ground-state (Fig. \ref{figurePP}). As the methanol lies in the same,
slightly offset, region of 6.035-GHz emission it too perhaps is in a
similarly high density regime. However of course if both the gas
density and temperature are high, then the collision rate will be
increased and both species of maser are likely to be quenched.
Combined with the 1612-MHz distribution, this would lead to the
assumption the excited-state OH traces the slightly cooler, dense gas,
the 6.668-GHz methanol the hotter dense gas and the 1612-MHz tracing
possibly the hottest dense gas (although the abundances of the
molecular species and the dust temperature may also vary).  The higher
density region could be indicative of a shock front, propagating away
from the UCHII.

Velocities of the maser features suggest a positional gradient, with 3
of the 4 methanol features showing a north-west to south-east
gradient, which is also seen in the excited-state OH at 6.035 GHz in 3
of the 5 features.  If a gradient does exist this would concur with
the ground-state OH picture and the possibility that the masers are
tracing an outflow or disk.  Interestingly two of the methanol
features and one of the 6.035-GHz OH features also demonstrate
tentative velocity gradients within their individual channel
components.  The two methanol features' internal gradients lie on a
north-east to south-west direction, i.e. perpendicular to the overall
velocity gradient across the features, and may suggest the masers are
tracing a planar shock (Elitzur, Hollenbach \& McKee 1992; Dodson,
Ojha \& Ellingsen 2004). The 6.035-GHz OH feature's internal gradient
meanwhile, lies parallel to the main distribution.

For comparison W3(OH) represents a star-formation region at a similar
distance to ON1, which has been studied at high resolution for
multiple transitions of masers (Menten et el. 1992; Sutton et
al. 2004; Wright et al. 2004; Harvey-Smith \& Cohen 2005; Etoka et
al. 2005; Harvey-Smith \& Cohen 2006).  The region has far more maser
features across the transitions and there is far more
``intermingling'' between the species and transitions, without the
possible separation that is seen in ON1. Within 15 mas in W3(OH) there
is a high percentage of associations between the excited-state OH and
the mainline ground-state OH (Etoka et al. 2005), which may not be the
case for ON1.  Both regions show a lack of association on the smallest
scales between 6-GHz OH and 1720-MHz OH sources, but unlike W3(OH),
ON1 demonstrates possible association of 6-GHz OH with 1612-MHz
OH. W3(OH) shows 27 per cent of 6.7-GHz methanol masers had associated
6-GHz OH maser emission, whilst in ON1 there is no association to 15
mas. Some of these differences could be accounted for by a difference
in the orientation of the regions and also the same caveat of internal
and external proper motions may affect the W3(OH) results (in so far
as the comparison of Etoka et al. 2005 compares data for mainline OH
observed in 1996, 4.7-GHz OH observed in 1993 and excited-state OH and
6.7-GHz methanol seen in 2001).

\subsection{Extended Emission}
The current observations did not detect any extended emission above
the 3$\sigma$ limit of $\sim$75 mJy beam$^{-1}$ at 100 mas resolution.
This contrasts with the detection in the W3(OH) star-formation region
(Harvey-Smith \& Cohen 2006) at a well established distance of 1.95
kpc (Xu et al. 2006).  If ON1 is at a similar distance, it lacks
diffuse methanol emission; however, if it is at the far-kinematic
distance, it might just be undetectable at present. Further study
would be required, both to determine whether extended emission is
prevalent in methanol maser sources, and thus expected to be seen, and
better determination of the distance to ON1 (perhaps through maser
astrometry as per Xu et al. 2006).

\subsection{Magnetic Field Strength \& Alignment} 
The average polarization angle of $-$60$^{\circ}$$\pm$28$^{\circ}$ is
consistent with the line of maser distribution, which is itself
perpendicular to the known H$^{13}$CO$^{+}$ outflow with a PA of
44$^{\circ}$ (Kumar et al. 2004). This suggests the emission could be
a propagating shock front, which agrees with the excited-state OH and
methanol lying in a potentially higher density region, as identified
in the previous Section. Faraday rotation is inversely proportional to
the square of the frequency and as such will strongly affect the
ground-state OH transitions. Nammahachak et al. (2006) estimate the
external rotation measure for ON1 to exceed $-$100 rad m$^{-2}$, more
than enough to disrupt any pattern present.  For the excited-state OH
at 6 GHz and methanol at 6.7 GHz internal Faraday rotation is
minimal. External Faraday rotation, which is calculated from the
standard Faraday rotation equation using typical values for
interstellar electron density and magnetic field as per Vlemmings et
al. (2006a), but adjusted for the distance of 1.8 kpc, is of the order
of 12$^{\circ}$ and 11$^{\circ}$ for the 6 GHz OH and 6.7 GHz methanol
respectively.

The coincident 6.031-GHz and 6.035-GHz Zeeman pairs (Z$_{1}$ and
Z$_{4}$) show the same field strength to within the errors, which concurs 
with the possibility of co-propagation mentioned previously. 
However, as this is the only coincidence
of the two transitions, and there is a similar sized field at
6.035-GHz for Z$_{2}$, the magnetic field strength may not be
intrinsically linked to the conditions necessary for maser
co-propagation.
The similarity of the 6.031-GHz and 6.035-GHz fields for Z$_{1}$ are
in contrast to that found by Desmurs et al. (1998), where the 6031-GHz
transition was seen to have stronger fields compared to the 6035 GHz
transition.

The tentative magnetic field strength of $-$18$\pm$6 mG derived from
methanol for the current observations is larger than that of both the
excited-state OH and the ground-state OH, although it is within the
established upper limit for the field strength derived through
methanol observations of W3(OH) of $-$22 mG (Vlemmings et al. 2006a).
The measured field strength implies the methanol may be tracing a
localised increase in density compared to the OH.  If the standard
scaling law of B$^{0.5}$ of Crutcher (1991) is applied, which has been
found to be valid up to the highest maser densities such as those
probed by H$_{2}$O masers (Vlemmings et al. 2006b), then the
implication is that the methanol masers occur in gas denser by a
factor of 5$-$10 compared to the OH masing gas at hydrogen densities
of $\sim$10$^{8}$.  However, caution must still prevail as the
magnetic field detection is only marginal.  If further study
demonstrated the presence of a stronger magnetic field in the regions
of methanol maser emission, this could well lead to a better
determination on the exact criteria for the 6.7 GHz emission or indeed
the star-formation stage it traces.

Comparison with the ground-state OH magnetic field strengths measured
by Nammahachak et al. (2006) highlight several similarities.  The
field strength of the 6.035-GHz OH seen towards the middle of the
linear distribution, that of Zeeman pair Z$_{2}$, is comparable to the
field strength of the nearby ($\sim$60 mas separation) 1665-MHz OH,
both giving 4.6 mG directed towards us. The the field detected at the
south-east end of the distribution, for the 6.035-GHz OH Zeeman pair
Z$_{4}$, is comparable to the 1720-MHz OH ($\sim$26 mas separation) of
1.0 mG. However, located between these two sets of similar fields,
the field strengths found for the methanol Zeeman splitting and the
6.035-GHz OH Zeeman pair Z$_{3}$ are both larger than the 1665-MHz OH
field (1.5 mG), although in this case there is a larger separation of
$\sim$190 mas.

\section[]{CONCLUSIONS}
New high-resolution MERLIN data demonstrate for the first time the
distribution and possible polarization properties of 6.7-GHz methanol
maser emission in the ON1 star-forming region. When combined with new
excited-state OH observations and existing ground-state OH data, and
correcting for the effects of Galactic motion, we see all transitions
lie in a similar region.  We see the structure of ON1 to be that of
two parallel, possibly offset, elongated distributions, one in
ground-state OH, one in interwoven excited-state OH and 6.7-GHz
methanol. The 6.031-GHz transition of excited-state OH shows a linear
polarization of 12 per cent, whilst the 6.035-GHz OH transition shows
linear polarization varying between 10 per cent and 19 per cent. In
comparison the methanol, as expected, demonstrates a lower value of
$\sim$1 per cent.  Consistent magnetic field strengths were observed
across the region for the excited-state OH, with a slight tendency for
smaller field strengths towards the south-east of the distribution,
which is in agreement with the known ground-state OH magnetic field
strengths. Zeeman splitting was detected for the first time in the
6.7-GHz methanol maser emission, demonstrating a possible
magnetic field strength of $-$18$\pm$6 mG.
 
A Zeeman pair at 6.031 GHz is seen in coincidence with a 6.035-GHz OH
Zeeman pair, with both having matching magnetic fields within the
errors.  This coincidence represented 100 per cent association to
$\sim$5 mas for the 6.031-GHz OH with the 6.035-GHz OH emission, but
only a 20 per cent association of 6.035 GHz with 6.031-GHz OH.  To the
same level of spatial separation there is no coincidence between the
methanol and excited-state OH transitions.

The observed interweaving of excited-state OH and methanol maser
features along the elongated distribution, together with the
separation of these masers from the ground-state OH is in agreement
with the postulation of Caswell (1997), that the two species delineate
similar or complimentary regions.  The separation of the individual
methanol and excited-state OH features on the scales afforded by high
resolution observations, could just be due to variations in the
relative abundances of the species, or it could be that the methanol
is tracing a slightly higher gas temperature or even denser regions
and thus a different component of the region surrounding the evolving massive star.  This is
complimented by the significantly higher magnetic field strength
suggested for the 6.7-GHz methanol maser.

Whether the maser features show a velocity gradient, and thus possibly
trace a disk, is not possible to judge as the number of features are
too few to draw statistically sound conclusions. However, the
consistent polarization angles and offset nature of denser gas imply
the masers trace a shock front, possibly in the form of a torus or
ring around a young stellar object.  This is also highlighted by
possible orthogonal velocity gradients across the individual
components of the methanol maser features.  The shock front hypothesis
concurs with the previous study of the ground-state transitions of OH
of Nammahachak et al. (2006).  The potential shock front lies
orthogonal to the known H$^{13}$CO$^{+}$ outflow and future proper
motion studies of the masers may be able to determine if they are
moving in synchronization with the outflow.

\section*{Acknowledgments}
JAG acknowledges the support of a Science and Technology Facilities Council (STFC) studentship.
WHTV was supported by a Marie Curie Intro-European fellowship within the 6th European Community 
Framework Program under contract number MEIF-CT-2005-010393.
JAG thanks L. Harvey-Smith for proposing the observations.
Figure 1 is reproduced from Szymczak et al. 2000 and JAG would like 
to thank the authors M. Szymczak, G. Hrynek and A. J. Kus for this. 
MERLIN is a national facility
operated by the University of Manchester on behalf of STFC. 
The authors would like to thank S. Ellingsen for his insightful comments and suggestions.
The authors would like to dedicate this paper to the memory of R. J. Cohen.

\label{lastpage}

\end{document}